\documentclass[slac_one]{revtex4}
\usepackage{graphicx}
\usepackage{fancyhdr}
\begin{document}

\newcommand{\dedx}{\mbox{${\rm d}E/{\rm d}x$}}

\title{Comparison of Geant4 hadron generators with data: 
a critical appraisal} %% Paper title goes here


\author{I. Boyko (for the 
HARP--CDP group\footnote{The members of the 
HARP--CDP group are:
A.~Bolshakova, 
I.~Boyko, 
G.~Chelkov, 
D.~Dedovitch, 
A.~Elagin, 
M.~Gostkin,
S.~Grishin,
A.~Guskov, 
Z.~Kroumchtein, 
Yu.~Nefedov, 
K.~Nikolaev and
A.~Zhemchugov 
from the Joint Institute for Nuclear Research, 
Dubna, Russian Federation;
F.~Dydak and  
J.~Wotschack 
from CERN, Geneva, Switzerland; 
A.~De~Min 
from the Politecnico di Milano and INFN, 
Sezione di Milano-Bicocca, Milan, Italy; and
V.~Ammosov, 
V.~Gapienko, 
V.~Koreshev, 
A.~Semak, 
Yu.~Sviridov, 
E.~Usenko and  
V.~Zaets 
from the Institute of High Energy Physics, Protvino, 
Russian Federation.} 
)}
\affiliation{Joint Institute for Nuclear Research, Dubna,
Russian Federation}

\begin{abstract}
Hadron generation models are indispensable for the
simulation and calibration of particle physics detectors.
The models used by the Geant4 simulation tool kit 
are compared with inclusive spectra of secondary protons 
and pions from the interactions with beryllium nuclei 
of $+8.9$~GeV/{\it c} protons and pions, 
and of $-8.0$~GeV/{\it c} pions. We report on 
significant disagreements between data and model 
predictions especially in the 
polar-angle distributions of secondary protons 
and pions. 
\end{abstract}

\maketitle

\thispagestyle{fancy}


\section{INTRODUCTION} % Section title should be in all capitals.

Precise cross-sections of secondary hadron production from the
interactions of protons and pions with nuclei are, {\it inter alia\/},
of importance for the improvement and physics validation of 
hadron generation models in Monte Carlo simulation tool kits 
such as Geant4~\cite{Geant4}. 

The HARP detector at the CERN PS took data in 2001 and 2002
with proton and pion beams with momentum from 
1.5 to 15~GeV/{\it c}, with a set of stationary targets ranging 
from hydrogen to lead, including beryllium. 
The detector combined a forward spectrometer with a 
large-angle spectrometer. The latter comprised a 
cylindrical Time Projection 
Chamber (TPC) around the target and an array of 
Resistive Plate Chambers (RPCs) that surrounded the 
TPC. The purpose of the TPC was track 
reconstruction and particle identification by \dedx . The 
purpose of the RPCs was to complement the 
particle identification by time of flight.

This paper reports comparisons of Geant4 predictions 
with data from the interactions with a 
5\% $\lambda_{\rm abs}$ beryllium 
target of $+8.9$~GeV/{\it c} protons and $\pi^+$'s, 
and of $-8.0$~GeV/{\it c} $\pi^-$'s. More details can be
found in Ref.~\cite{GEANTpub}.

\section{DETECTOR CHARACTERISTICS AND PERFORMANCE}

% tables should appear as floats within the text

\begin{figure*}[h]
\begin{center}
\begin{tabular}{cc} 
\includegraphics[height=0.3\textwidth]{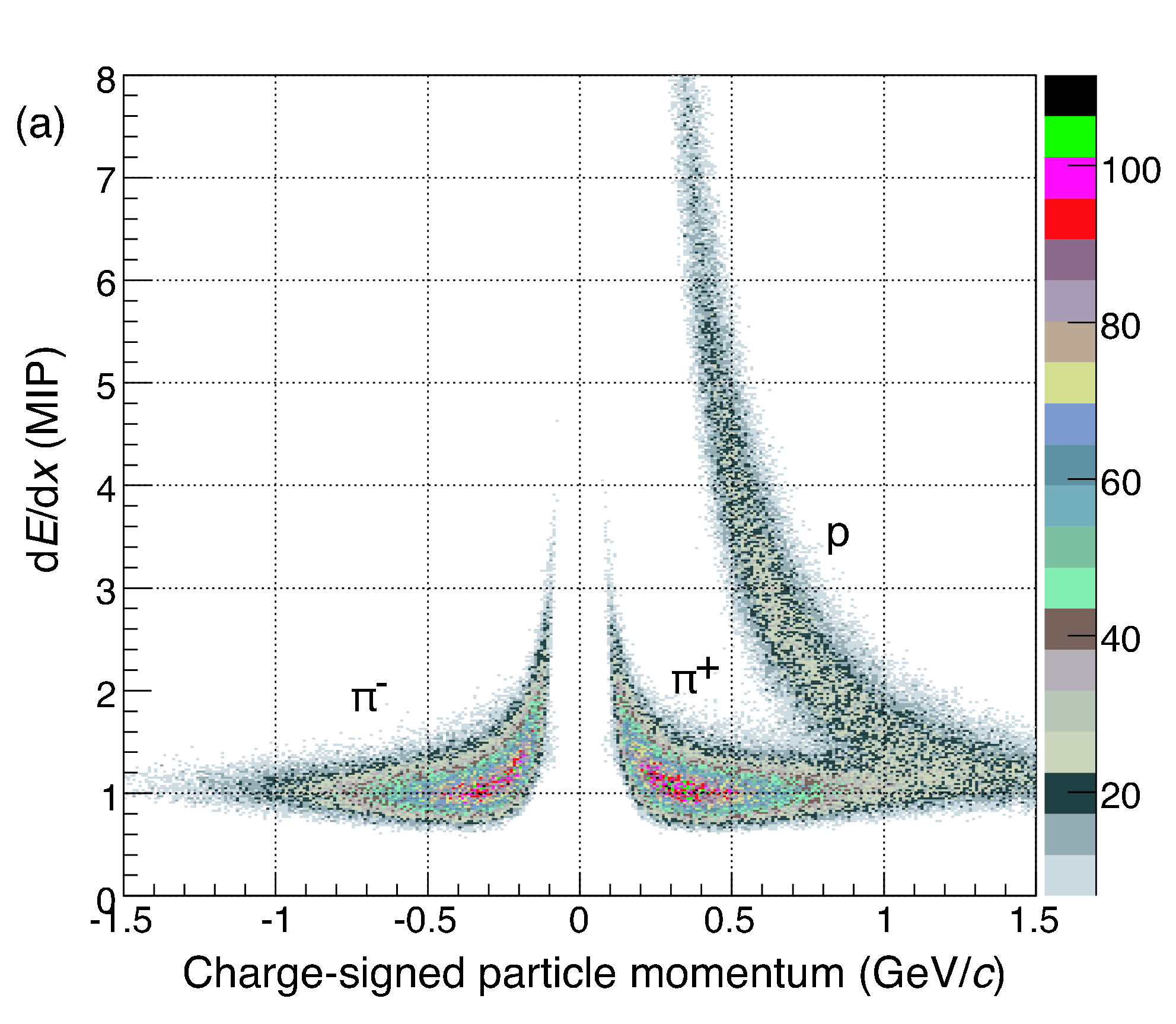} &
\includegraphics[height=0.3\textwidth]{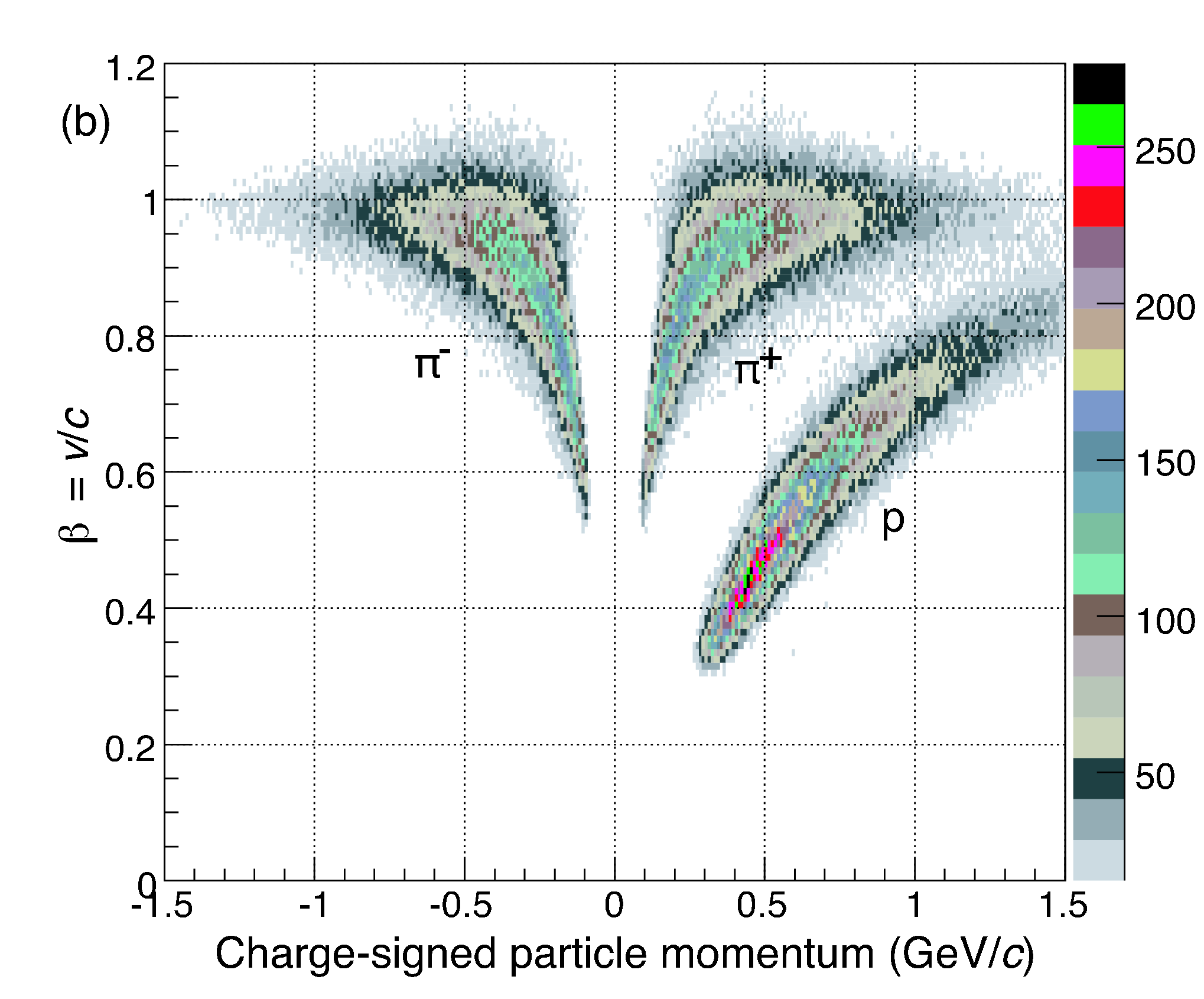} \\
\end{tabular}
\end{center}
\caption{Specific ionization d$E$/d$x$ (left panel) and velocity $\beta$
(right panel) versus the charge-signed momentum of positive and 
negative tracks 
in $+8.9$~GeV/{\it c} data.}
\label{dedxandbeta}
\end{figure*}

For the work reported here, only the HARP large-angle spectrometer
was used~\cite{TPCpub,RPCpub}. Its salient technical 
characteristics are stated in Table~\ref{LAcharacteristics}. The 
good particle identification 
capability stemming from \dedx\ in the TPC and from time of flight 
in the RPC's is demonstrated in Fig.~\ref{dedxandbeta}. 
\begin{table}[ht]
\vspace*{2mm}
\begin{center}
\caption{Technical characteristics of the HARP large-angle spectrometer}
\begin{tabular}{|c|c|}
\hline 
\textbf{TPC} & \textbf{RPCs}  \\
\hline
\hline
$\sigma(1/p_{\rm T}) \sim 0.20-0.25$~(GeV/{\it c})$^{-1}$ &
   Intrinsic efficiency $\sim 98$\%  \\
$\sigma(\theta) \sim 9$~mrad  & $\sigma$(TOF) $\sim 175$~ps  \\
$\sigma({\rm d}E/{\rm d}x) / {\rm d}E{\rm d}x \sim 0.16$  &  \\ 
\hline
\end{tabular}
\label{LAcharacteristics}
\end{center}
\end{table}

\section{Geant4 PHYSICS LISTS}

The Geant4 simulation tool kit provides several 
physics models of hadronic interactions of hadrons with nuclei, 
and collections of such models, termed `physics lists'.  

In the so-called `low-energy' domain
(defined as kinetic energy 
$E$ of the incoming hadron below some $25$~GeV), a modified version
of the GHEISHA package of Geant3 is used in many physics lists: 
the Parameterized
Low-Energy Model 
(`LE\_GHEISHA'). Optionally, 
for $E$ below a few GeV, the 
Bertini Cascade~\cite{bertini} (`BERT') or the 
Binary Cascade~\cite{binary} (`BIC') 
models can be enabled, with a view to simulating the cascading
of final-state hadrons when they move through nuclear matter.  
As an alternative to LE\_GHEISHA, a modified 
version of the FRITIOF string fragmentation 
model~\cite{fritiof} (`FTF') is available. 

In the so-called `high-energy' domain, mostly the Quark--Gluon String Model
(`QGSM') is used, with FTF and the Parameterized High-Energy Model
(`HE\_GHEISHA') as alternatives. Further terms that are explained in 
Ref.~\cite{Geant4RefMan}, are `PRECO' for the 
Pre-compound model, `QEL' for the Quasi-elastic scattering model,
and `CHIPS' for the Chiral Invariant Phase Space model.

\begin{figure*}[ht]
\begin{center}
\begin{tabular}{ccc} 
\includegraphics[width=0.3\textwidth]{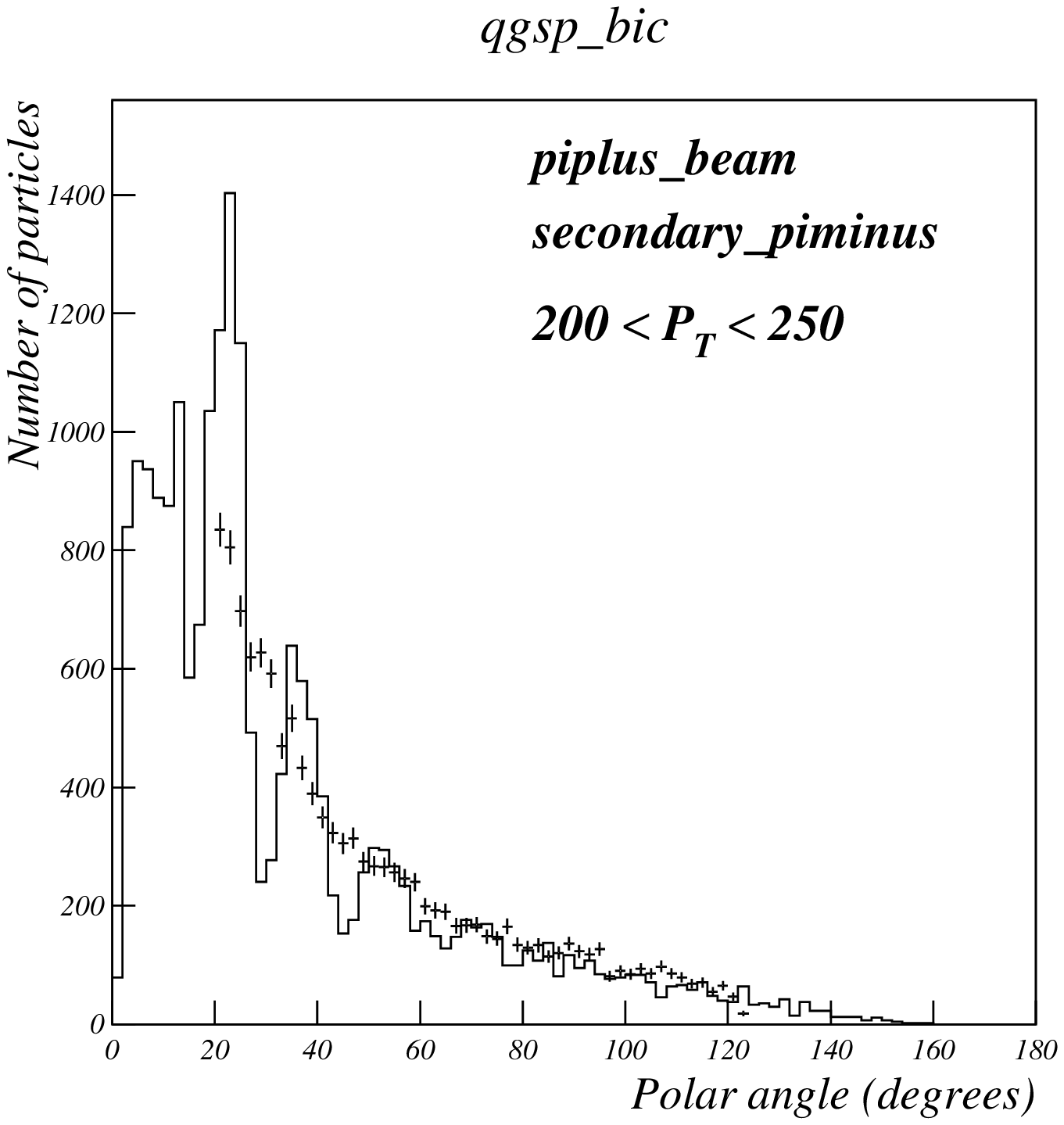} & 
\includegraphics[width=0.3\textwidth]{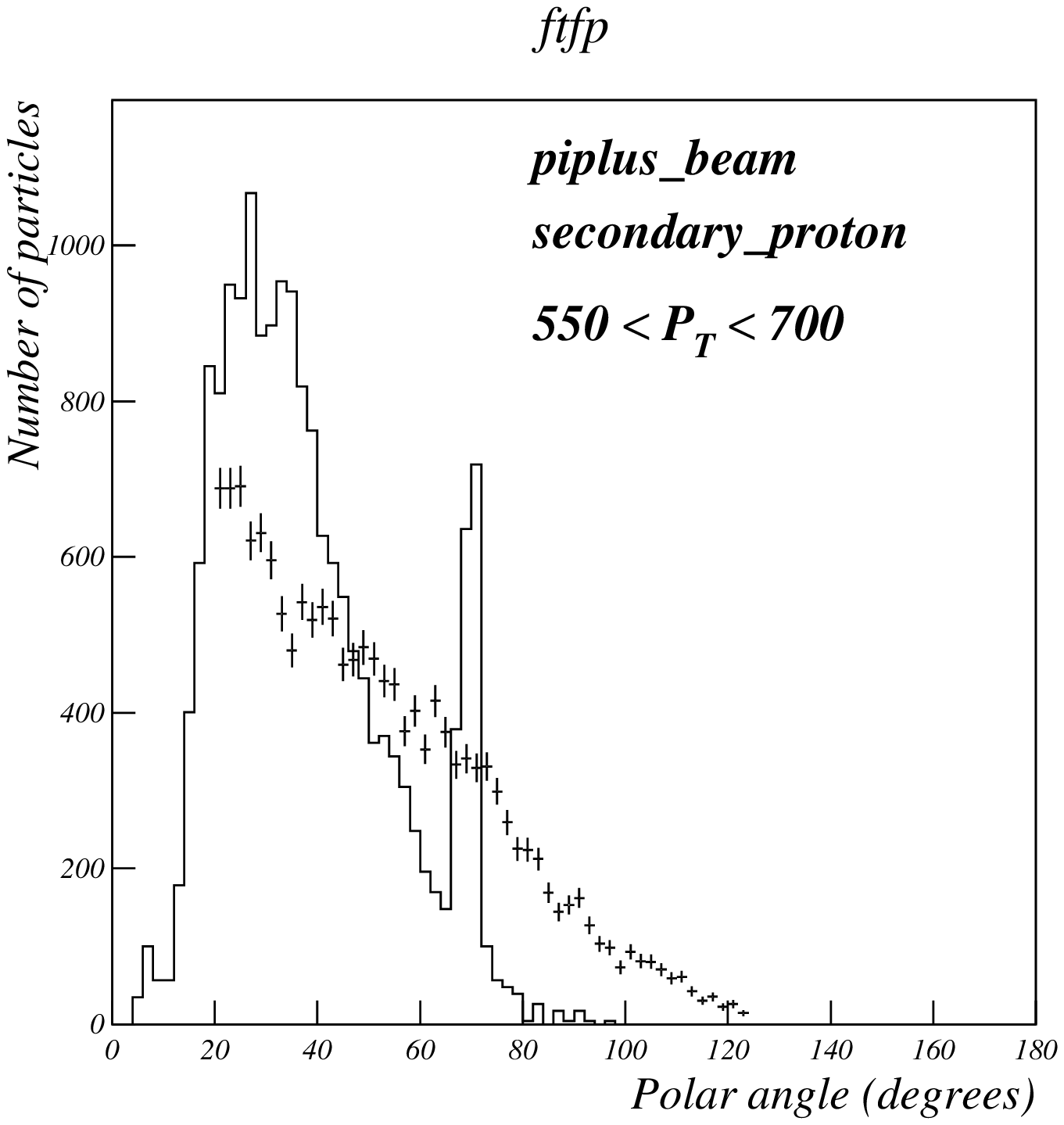} &
\includegraphics[width=0.3\textwidth]{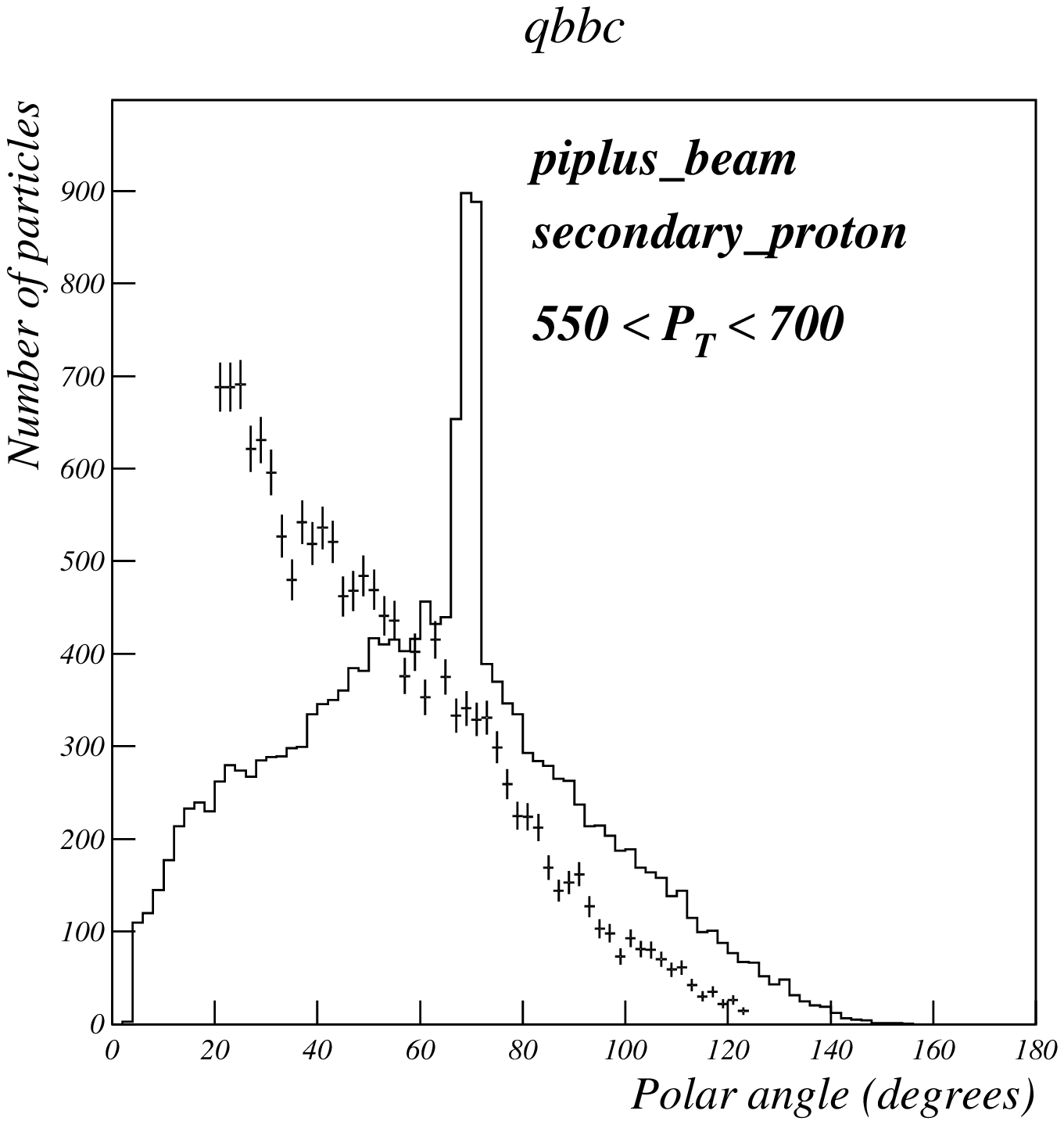} \\
\end{tabular}
\end{center}
\caption{QGSP\_BIC physics list (left panel): polar-angle distributions of 
$\pi^-$'s for incoming $\pi^+$'s;
FTFP physics list (middle panel): polar-angle distributions of protons
for incoming $\pi^+$'s;
QBBC physics list (right panel): polar-angle distributions of protons
for incoming $\pi^+$'s; crosses denote data, full lines the Geant4 
simulation.}
\label{compphyslist}
\end{figure*}

Figure~\ref{compphyslist} shows comparisons of polar-angle
distributions with data for three representative Geant4 physics 
lists: QGSP\_BIC, FTFP, and QBBC. There are serious disagreements between 
data and model predictions.

\section{HARP--CDP PHYSICS LIST}

\begin{figure*}[h]
\vspace*{8mm}
\begin{center}
\begin{tabular}{cc}  
\includegraphics[width=0.3\textwidth]{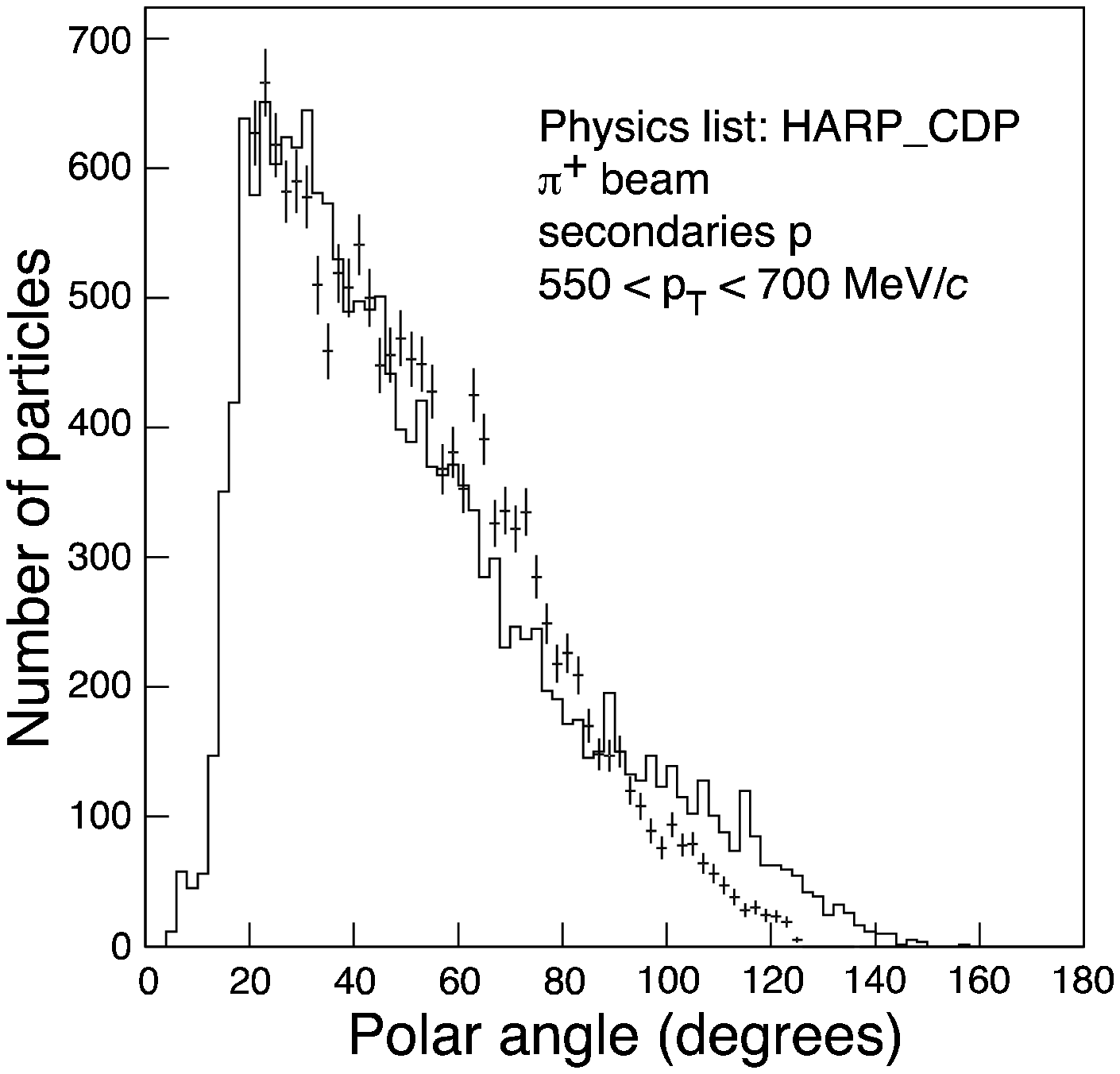} &
\includegraphics[width=0.3\textwidth]{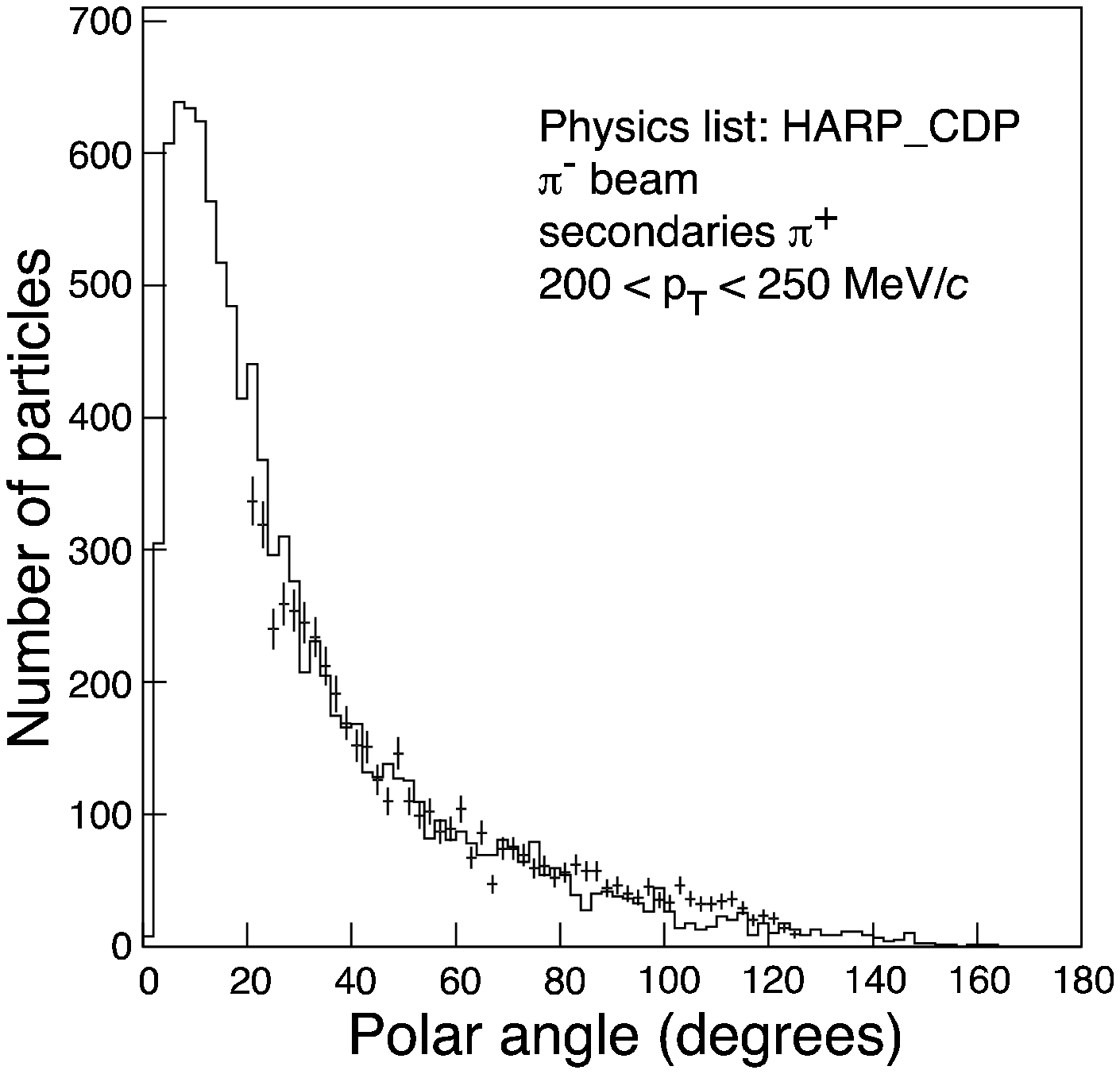} \\ 
\end{tabular}
\end{center}
\caption{HARP\_CDP physics list: polar-angle distributions of protons
for incoming $\pi^+$'s (left panel), and of $\pi^+$'s for incoming $\pi^-$'s
(right panel); crosses denote data, full lines the Geant4 
simulation.}
\label{compharpcdp}
\end{figure*}

For the determination of hadronic cross-sections, we have used for 
incoming beam protons below $\sim$10~GeV/{\it c}
the QGSP\_BIC physics list and see no strong reason to reconsider 
this choice. For incoming beam pions and for protons 
above $\sim$10~GeV/{\it c}, none of the standard 
physics lists for hadronic interactions was acceptable, so we 
had to build our private HARP\_CDP physics list. 
This physics list starts from the QBBC physics list. Yet  
the Quark--Gluon String Model is replaced by the 
FRITIOF string fragmentation model for
kinetic energy $E>6$~GeV; for $E<6$~GeV, the Bertini 
Cascade is used for pion interactions, and the Binary Cascade for 
proton interactions; 
elastic and quasi-elastic scattering is disabled.
Figure~\ref{compharpcdp} shows the comparison of data 
with the simulation results from the HARP\_CDP physics list.
The agreement is good and permits its use, after due weighting,
in the analysis of the interactions of few GeV/{\it c} protons and 
pions. 

\end{document}